\documentclass[conference]{IEEEtran}
\IEEEoverridecommandlockouts
\usepackage{cite}
\usepackage{amsmath,amssymb,amsfonts}
\usepackage{algorithmic}
\usepackage{graphicx}
\usepackage{textcomp}
\usepackage{xcolor}
\usepackage{booktabs}
\usepackage{url}
\def\BibTeX{{\rm B\kern-.05em{\sc i\kern-.025em b}\kern-.08em
    T\kern-.1667em\lower.7ex\hbox{E}\kern-.125emX}}
\begin{document}

\title{Prompt-Unseen-Emotion: Zero-shot Expressive Speech Synthesis with Prompt-LLM Contextual Knowledge for Mixed Emotions\\
\author{\IEEEauthorblockN{Xiaoxue Gao, Huayun Zhang and Nancy F. Chen}\\
\textit{Institute for Infocomm Research (I2R), Agency for Science, Technology, and Research (A*STAR), Singapore}\\
{Gao\textunderscore Xiaoxue@i2r.a-star.edu.sg, Zhang\textunderscore Huayun@i2r.a-star.edu.sg, nfychen@i2r.a-star.edu.sg}}
}

\maketitle

\begin{abstract}
Existing expressive text-to-speech (TTS) systems primarily model a limited set of categorical emotions, whereas human conversations extend far beyond these predefined emotions, making it essential to explore more diverse emotional speech generation for more natural interactions. To bridge this gap, this paper proposes a novel prompt-unseen-emotion (\textit{PUE}) approach to generate unseen emotional speech via emotion-guided prompt learning.
\textit{PUE} is trained utilizing an LLM-TTS architecture to ensure emotional consistency between categorical emotion-relevant prompts and emotional speech, allowing the model to quantitatively capture different emotion weightings per utterance. During inference, mixed emotional speech can be generated by flexibly adjusting emotion proportions and leveraging LLM contextual knowledge, enabling the model to quantify different emotional styles.
Our proposed \textit{PUE} successfully facilitates expressive speech synthesis of unseen emotions in a zero-shot setting. 
\end{abstract}


\section{Introduction}
Humans naturally produce speech with a wide range of emotional variations \cite{yasuda2023text,chen2023vector,khanam2022text,nose2007style,seamless2023}. Expressive speech synthesis seeks to capture this diversity by converting text into speech that not only sounds human but also conveys a genuine emotional tone~\cite{um2020emotional,diatlova2023emospeech,lee2017emotional,li2024mm,guo2023emodiff,yang2024instructtts}. To generate genuinely realistic expressive speech, expressive text-to-speech (TTS) systems must account for a range of factors that extend beyond the textual content alone~\cite{luo2019emotional,cai2021emotion,wu2024laugh,kim2021expressive,li2021controllable}.

These include subtle ways in which multiple emotions are expressed through the intricate interplay of human emotional characteristics and the robustness of the model \cite{ma2023emotion2vec,gao2025ttslow,inoue2024hierarchical}, especially given that humans are capable of experiencing approximately 34,000 distinct emotions and even multiple emotional states simultaneously \cite{plutchik2001nature,braniecka2014mixed,tang2023emomix}. For instance, humans naturally feel mixed emotions in their daily life, such as attending a child's wedding can evoke both happiness seeing their child embark on a new journey and a poignant sense of loss~\cite{zhou2022speech}. This underscores the importance of emotional TTS models that can accurately simulate the co-occurrence of diverse emotional states to reflect human experiences~\cite{zhou2022speech}.

Expressive speech synthesis has been extensively studied relying on the data-driven training approaches, including Tacotron \cite{li2021controllable,um2020emotional,li2024mm,wang2018style}, long-short-term memory network~\cite{lei2021fine,liu2024emotion}, FastSpeech architecture~\cite{kim2021expressive,diatlova2023emospeech,lee2017emotional,li2024mm,yang2024instructtts}, VITS architecture~\cite{zhao2023emotion}, diffusion models \cite{guo2023emodiff}, flow-matching models~\cite{wu2024laugh} and large language model (LLM)-based systems \cite{zhou2024emotional,gao2024emo,ye2025llasa,gao2025emo}.
These data-driven approaches rely on learning the temporal emotional structure of a fixed set of speech emotions, typically limited to at most eight categories in existing emotional TTS databases. As a result, their expressiveness is constrained, as the synthesized speech is restricted to reproducing only the emotion types present in the training data.

Although these methods excel in generating emotional speech for specific categories such as anger, happiness, sadness, and surprise, the generation of unseen emotional expressions, such as disappointment, remains an open challenge~\cite{wu2024laugh,zhao2023emotion,lei2021fine,liu2024emotion}. Existing methods are still struggling to capture complex unseen emotional styles needed for real-world applications~\cite{li2021controllable,um2020emotional,li2024mm,wang2018style}. One particularly challenging aspect is generating mixed-emotion speech, which requires modeling intricate combinations of emotional expressions.

In line with this, the emotion wheel theory proposes that all emotions stem from eight primary emotions: anger, fear, sadness, disgust, surprise, anticipation, trust, and happiness~\cite{plutchik2013theories}. The remaining 34,000 emotional states are viewed as mixed or derivative forms of these foundational categories~\cite{plutchik2013theories}. For instance, disappointment can be conceptualized as a blend of surprise and sadness~\cite{cross2016changing,zhou2022speech}.
The only prior work for mixed emotional TTS employs a VITS-based framework with a relative scheme~\cite{zhou2022speech}, which, although constrained in speech quality, provides valuable insights for this study.
In contrast, other mixed-emotion generation methods focus on emotional voice conversion~\cite{zhou2022mixed}, which takes speech as input rather than text, making it fundamentally different from emotional TTS while still providing relevant insights for this work.

\vspace{-0.2cm}
\begin{figure*}[t!]
\centering
\includegraphics[width=172mm]{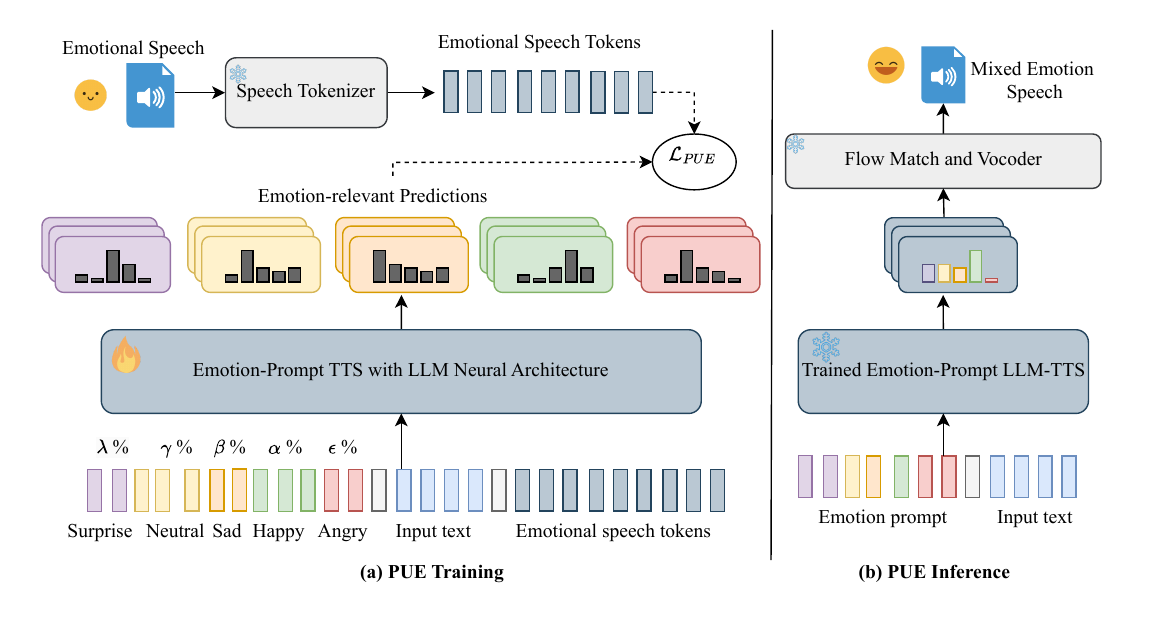}
\vspace{-0.6cm}
\caption{An overview of the proposed \textit{PUE} approach: (a) the \textit{PUE} training process and (b) the \textit{PUE} inference process.}
\vspace{-0.4cm}
\label{overall}
\end{figure*}

Motivated by the success of in-context learning abilities in LLMs \cite{achiam2023gpt,team2023gemini} and the need to model more diverse emotional styles that go beyond predefined emotional categories~\cite{zhou2022speech}, we propose a prompt-unseen-emotion (\textit{PUE}) approach with emotion-guided prompt learning technique to capture the nuanced characteristics of mixed emotional expressions. Emotion-guided prompt learning is designed to capture and quantify different emotional styles by modeling the emotional proportions embedded within the prompts. \textit{PUE} is trained using an LLM-based TTS architecture, which conditions the emotional context on speech generation to maintain emotional consistency between the contextual prompts and the corresponding speech data. By incorporating various emotion style descriptions into the emotion-relevant contexts, the \textit{PUE} leverages the LLM's in-context learning capabilities to learn the distribution of different emotional proportions for each utterance. This enables the generation of unseen emotions by adjusting the emotion proportions in the prompts during inference, offering flexibility in emotional expression.

The main contributions of this paper are as follows:
\begin{itemize}
    \item The proposed \textit{PUE} method is the first to leverage the in-context learning capabilities of LLMs through prompting for the purpose of expressive speech generation of unseen emotion types that are challenging to obtain in real-world datasets.
    \item  \textit{PUE} facilitates zero-shot expressive speech synthesis by simply modifying emotion-relevant prompts during inference, enabling mixed-emotion speech synthesis without the need for additional training on new emotion categories. 
    \item Extensive objective and subjective evaluations demonstrate that \textit{PUE} outperforms baseline methods, achieving superior mixed-emotion speech generation. 
\end{itemize}

\section{Methodology-PUE}
We propose the \textit{PUE} method for expressive speech synthesis, which incorporates emotion-guided prompt learning within a large language model (LLM)-based TTS architecture, as depicted in Fig.~\ref{overall}.

\subsection{PUE Overview}
We introduce \textit{PUE}, an expressive TTS approach that generates emotional speech conditioned on both textual content and specified emotional inputs, with a particular focus on synthesizing mixed emotions that are not encountered during training. This method combines emotion-guided prompts with an LLM-based TTS network, optimized to maximize the likelihood of producing a speech token sequence that accurately reflects the emotional cues provided by the emotion-guided instructional data, as illustrated in Fig.~\ref{overall} (a). During inference, \textit{PUE} synthesizes mixed-emotion speech by adjusting emotion proportions in the input prompts, which are then processed by a frozen flow-matching model and followed by a fixed vocoder to generate unseen expressive speech, as shown in Fig.~\ref{overall} (b).

\subsection{Emotion-guided Prompt Learning}
We propose leveraging emotion-guided prompt learning with an LLM-TTS neural architecture $\theta$ to take advantage of the instruction-following abilities and in-context learning capabilities from LLMs using paired emotional textual and speech data. The training data is structured with the emotion-guided prompt alongside the input text as follows:

\begin{equation}
        d_i \in D_\texttt{e} = EP \texttt{<EOP>} t_i \texttt{<T>} s_i \texttt{<E>}
\end{equation}

where $EP$, $t_i$, $s_i$, $\texttt{<EOP>}$, and $\texttt{<T>}$ represent the emotion-guided prompts, the textual token sequence, the emotional speech token sequence corresponding to the current emotion, the-end-of-prompt token marking the end of emotion-guided knowledge trigger, the turning token from text to speech, and the-end-of-sequence token, respectively.

To enable mixed-emotion contextual prompt learning, we propose conditioning the LLM-TTS model with diverse emotion styles by formulating the emotion-guided prompt $EP$ below:
\begin{equation}
\begin{split}
        EP = A\texttt{ }man/woman\texttt{ }speaks\texttt{ }an utterance\texttt{ }with,\\
        \texttt{ }\alpha\texttt{ }percent
        \texttt{ }happy\texttt{ }emotion,\\ 
        \beta\texttt{ }percent\texttt{ }sad\texttt{ }emotion,\\
        \gamma\texttt{ }percent\texttt{ }neutral\texttt{ }emotion,\\
        \epsilon\texttt{ }percent\texttt{ }angry\texttt{ }emotion,\\ 
        \texttt{ }and\texttt{ }\lambda\texttt{ }percent\texttt{ }surprise\texttt{ }emotion.
\end{split}
\end{equation}
In the proposed emotion-guided prompt module $EP$, we integrate five canonical emotion representations corresponding to the primary emotion categories available in the dataset, enabling automatic emotion-conditioned TTS synthesis. The conditional TTS architecture contains a speech tokenizer and an emotion-prompted LLM-TTS model, which includes a text encoder and an auto-regressive LLM decoder. The speech tokenizer encodes the acoustic input into discrete emotional token sequences, while the auto-regressive LLM decoder predicts the emotion-relevant probability distribution of emotional speech tokens across the diverse emotion styles.

Emotion-guided prompts are automatically constructed by assigning different scalars to the emotion-relevance parameters $\alpha$, $\beta$, $\gamma$, $\epsilon$, and $\lambda$, based on the emotional category of the input sample in the dataset. For each training sample, a distinct emotion-relevance prompt is instantiated based on its labeled emotion category: the parameter corresponding to the target emotion is set to 100, while the remaining parameters are set to 0. 
Although the dataset lacks mixed-emotion speech samples, the proposed prompt design allows for the flexible specification of arbitrary combinations of emotion-relevance parameters. 

This facilitates the implicit modeling of mixed-emotion generation through the instruction-following capabilities inherent to large language models. 
Through this mechanism, the proposed \textit{PUE} is contextually guided with the weightings of various emotion proportions per utterance in a quantitative manner, allowing finer control over the emotional composition of each utterance. Given the high cost of collecting and annotating mixed-emotion speech data, our \textit{PUE} offers substantial flexibility for zero-shot expressive speech generation without requiring such data during training.

\subsection{\textit{PUE} Training Objective}
Motivated by the success of label smoothing Kullback-Leibler loss in speech generation~\cite{du2024cosyvoice}, we employ an emotion-guided Kullback-Leibler (KL) loss to minimize the divergence between the emotional probability distribution predicted by $\theta$, denoted as $P_{\theta}$, and the target emotional distribution $P$:
\begin{equation}
\begin{split}
\mathcal{L}_{PUE} = KL(P_{\theta}||P),\\
KL(P_{\theta}||P) = \mathbb{E}_{d_i\sim D_\texttt{e}} \left[ p(s_i | EP, t_i) \log \frac{p(s_i | EP, t_i)}{p_{\theta}(s_i | EP, t_i)} \right]
\end{split}
\end{equation}
In this manner, $\theta$ learns to produce the mixed-emotion speech token sequences that are consistent with the emotion-relevant contexts provided in the input instruction, ensuring that the generated expressive speech captures the weighting of emotion mixture contexts as indicated by $EP$. 

During inference, emotion-guided prompts can be adjusted by assigning different values to $\alpha$, $\beta$, $\gamma$, $\epsilon$ and $\lambda$, which define the proportions of relative emotion styles, thus facilitating the generation of mixed-emotion speech.

\section{Experiments}
In this section, we outline the setups of the experiments and datasets, along with the design of both subjective and objective evaluations.
\subsection{Experimental Setup}
We employ Cosyvoice-300M-Instruct model (cosy)~\cite{du2024cosyvoice} and the Mixed Emotion model (mix)~\cite{zhou2022speech} as strong baselines.
For both cosy baseline and our proposed \textit{PUE} model, a pretrained flow-matching model and a pretrained HiFi-GAN vocoder~\cite{du2024cosyvoice} are used during inference. Man and woman in Equation (2) are decided by the gender of the desired speech output. \textit{PUE} is trained for one epoch using dynamic batching on four GPUs, whose LLM-based TTS model, the speech tokenizer, and the text encoder are initialized from CosyVoice, maintaining the same architectural configurations.

Following \cite{zhou2022speech} during inference, we generate zero-shot mixed-emotional speech by combining a primary emotion (surprise) with three other emotions (happy, angry, and sad) by setting $\alpha$ as 30/60/90, setting $\beta$ as 30/60/90, setting $\gamma$ as 0, setting $\epsilon$ as 30/60/90, and setting $\lambda$ as 100, resulting in the mixed emotions of delight (surprise + happy), outrage (surprise + angry), and disappointment (surprise + sad), respectively. These mixed emotions are considered more easily perceivable by listeners, as demonstrated in psychological studies~\cite{braniecka2014mixed,zhou2022speech,plutchik2013theories}.

\subsection{Datasets}
To compare with the mix baseline \cite{zhou2022speech}, we use the same two English speakers (one female: 0019 and one male: 0013) from the ESD dataset \cite{zhou2022emotional} for experiments, encompassing five distinct emotional categories: neutral, angry, happy, sad and surprise. Each speaker contributes 350 utterances for each emotion, resulting in approximately 1,750 utterances and 1.2 hours of speech per speaker. We adopt the official train/validation/test splits provided in \cite{zhou2022emotional} and follow the mix baseline configuration from~\cite{zhou2022speech}, where the validation and test sets contain 30 and 20 utterances per emotion, respectively. As the dataset contains only single-emotion labels, one parameter from emotion-relevance parameters $\alpha$, $\beta$, $\gamma$, $\epsilon$, and $\lambda$ corresponding to the target emotion is set to 100, while the others are set to 0. For example, if the target emotion is happy, then $\alpha$ is set to 100, and the remaining parameters are set to 0.

\subsection{Subjective Evaluations}
We conduct extensive subjective evaluations to compare the proposed \textit{PUE} with baselines. 18 listeners participate all subjective evaluation tests.
\subsubsection{AB Preference Tests}
Listeners are asked to choose the better one between samples from two systems (A and B) based on speech quality. Two AB preference tests are conducted: cosy vs. \textit{PUE} and mix vs. \textit{PUE}, each using 24 balanced mixed-emotion samples, which covers both original test data with single emotion samples and mixed-emotion samples. In mixed samples, surprise is combined with varying proportions of angry, sad, and happy emotions (e.g., surprise + 30\% angry, surprise + 60\% angry, and surprise + 90\% angry).

\subsubsection{Best-worst Scaling (BWS) Tests}
To investigate how humans perceive mixed emotions in the proposed \textit{PUE}, we conduct best-worst scaling tests by blending 100 \% surprise with anger at varying levels of angry prompt (0 \%, 30 \%, 60 \% and 90 \%). Listeners are tasked with selecting the best and worst samples based on their perception of the intended emotional style.

\subsubsection{Mean Opinion Score (MOS) Tests}
To compare speech quality and mixed-emotion effects between the proposed \textit{PUE} and the mix and cosy baselines, listeners are asked to rate overall speech quality on a scale of 1 (bad), 2 (poor), 3 (fair), 4 (good) and 5 (excellent) for the mixed-emotion generation of outrage, disappointment, and delight. Besides, MOS tests are also conducted to compare the speech quality and mixed emotion perception of proposed PUE under different test cases: original test database; 100\% surprise mixed with varying proportions of other emotions: 30\%, 60\%, and 90\% for angry, sad, or happy.

\subsection{Objective Evaluations}
To assess the intelligibility of generated emotional speech, we apply Whisper-Large-v3 \cite{radford2022whisper} on the speech samples to recognize the text and calculate the word-error-rate (WER) for both baselines and the proposed \textit{PUE} across different test cases: (1) the original test dataset with single emotions; (2) surprise mixed with 30 \% angry, happy or sad; (3) surprise mixed with 60 \% angry, happy or sad and (4) surprise mixed with 90 \% angry, happy, or sad.

\begin{table}
\centering
\caption{Objective evaluation results comparison of the proposed \textit{PUE} with baselines on speech intelligibility in terms of speech intelligence through word error rate (WER).}
\begin{tabular}{l|ccc}
\toprule
\textbf{TTS Models} & \textbf{mix}\cite{zhou2022speech} & \textbf{cosy}\cite{du2024cosyvoice} & \textbf{\textit{PUE}} \\\midrule
original test  & 32.59& 51.85& \textbf{26.67}
\\
surprise + 30\% others & 32.59& 47.41& \textbf{25.19}
\\
surprise + 60\% others & 31.11& 42.22& \textbf{25.93}
\\
surprise + 90\% others & 27.41& 39.26& \textbf{25.93}\\
\bottomrule
\end{tabular}
\label{wer}
\end{table}

\begin{table}
\centering
\caption{BWS result comparison of the proposed \textit{PUE} approach on mixing 100 \% surprise with anger at varying levels of angry (0\%, 30\%, 60\% and 90\%).}
\begin{tabular}{lrr}
\toprule
\textbf{Outrage: perception of angry} & \textbf{Best (\%)} & \textbf{Worst (\%)} \\\midrule
Mix surprise with 0\% angry & 5.56 & \textbf{80.56} \\
Mix surprise with 30\% angry & 0 & 2.78 \\
Mix surprise with 60\% angry & 0 & 11.11 \\
Mix surprise with 90\% angry & \textbf{94.44} & 5.56 \\
\bottomrule
\end{tabular}
\label{BWS}
\end{table}

\begin{figure}[t]
\centering
\vspace{-0.2cm}
\includegraphics[width=75mm]{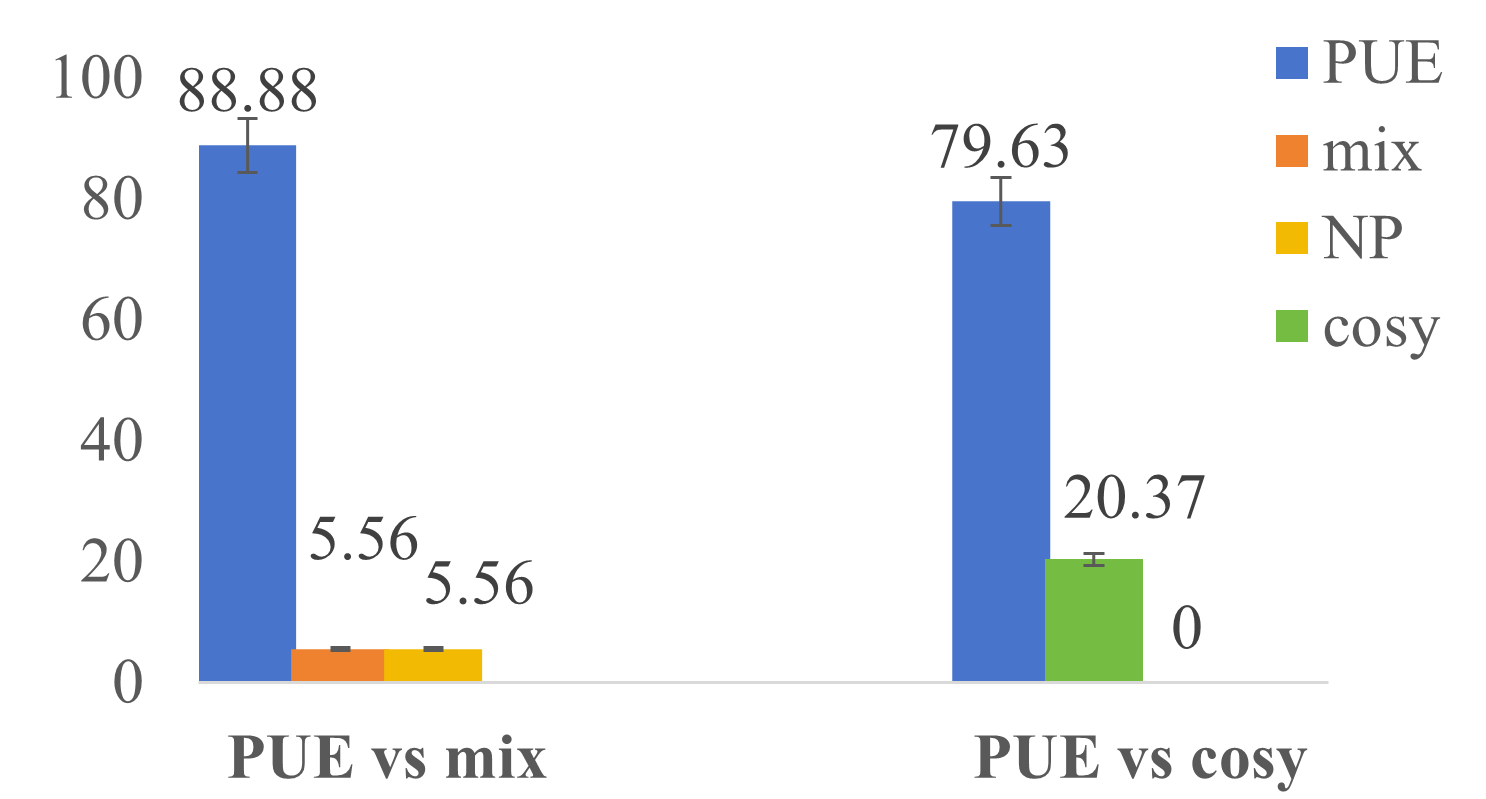}
\caption{Comparison of subjective evaluation results for AB preference test results with 95\% confidence interval between mix baseline and the proposed \textit{PUE} model, and between cosy baseline and the proposed \textit{PUE} model on seen single emotional speech samples. NP indicates no preference.}
\label{singleAB}
\vspace{-0.2cm}
\end{figure}

\begin{figure}[t]
\centering
\includegraphics[width=77mm]{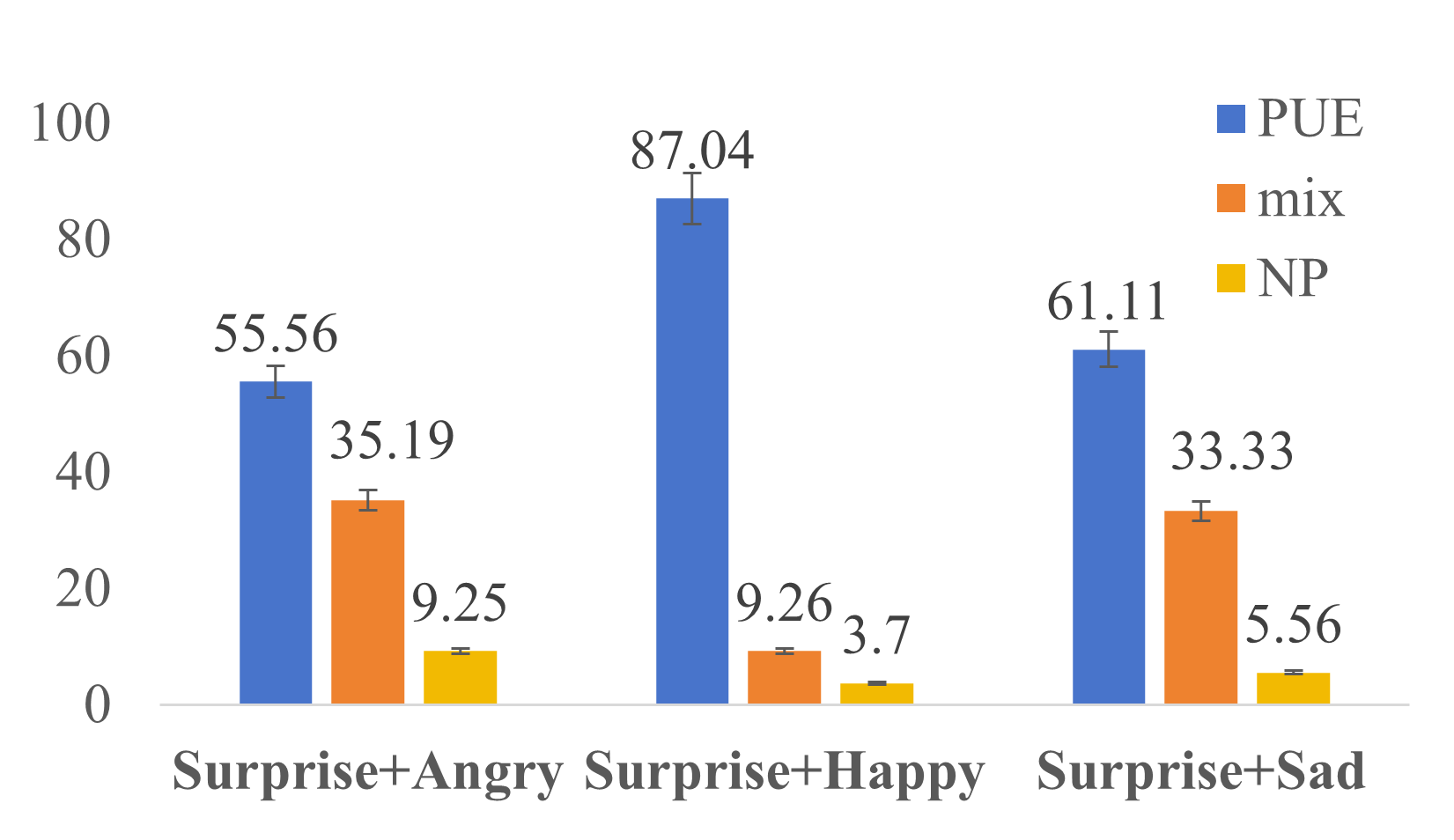}
\caption{Comparison of AB preference test results with 95\% confidence interval between the mix baseline and the proposed \textit{PUE} models, examining emotion expression of angry, sad, and happy, with 100\% surprise mixed in varying proportions: 30\%, 60\%, and 90\% for angry, sad, and happy. NP indicates no preference.}
\label{ABmix}
\vspace{-0.2cm}
\end{figure}

\begin{figure}[t]
\centering
\includegraphics[width=77mm]{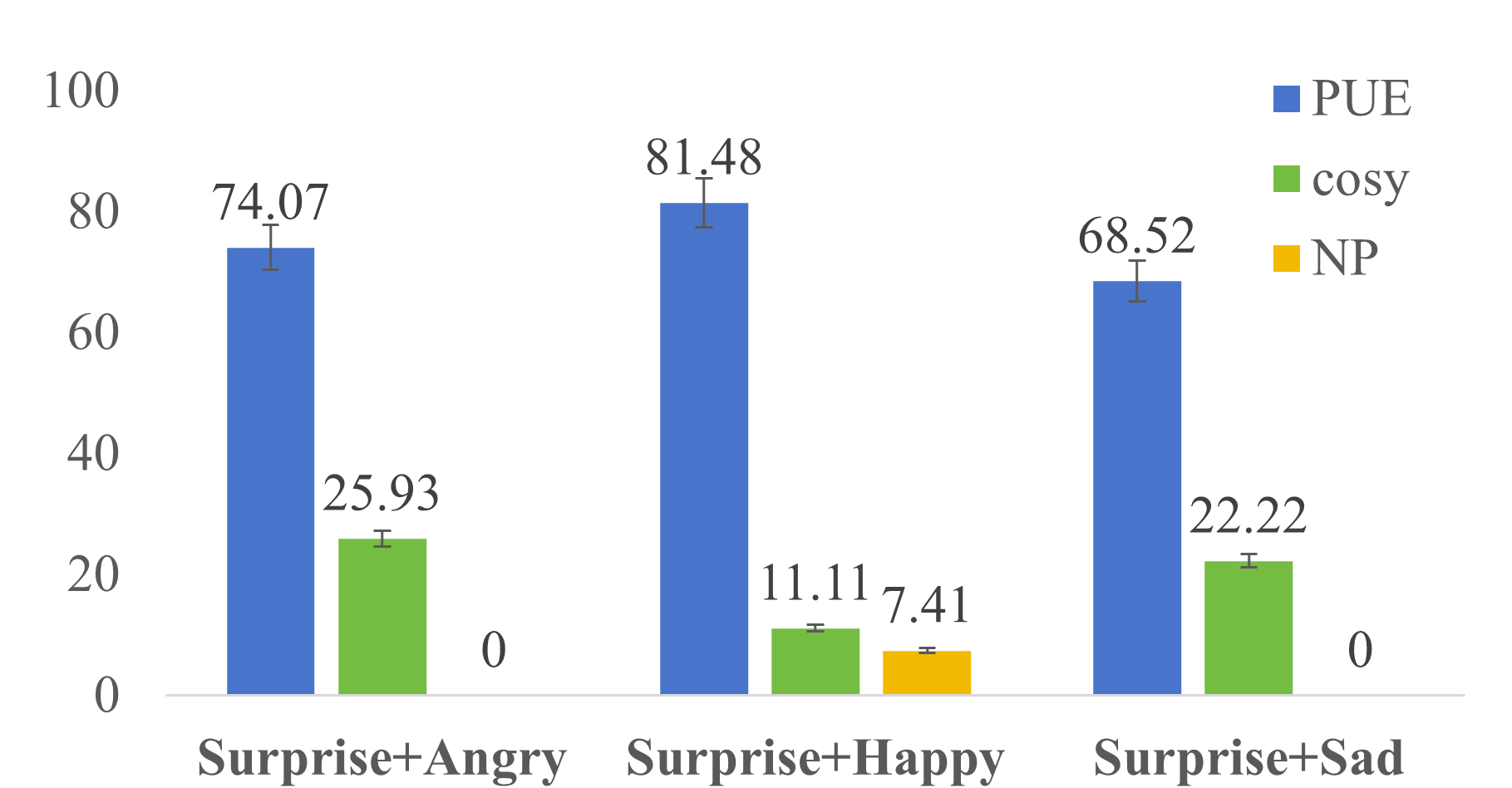}
\caption{Comparison of AB preference test results with 95\% confidence interval between the cosy baseline and the proposed \textit{PUE} models, examining emotion expression of angry, sad, and happy, with 100\% surprise mixed in varying proportions: 30\%, 60\%, and 90\% for angry, sad, and happy. NP indicates no preference.}
\vspace{-0.2cm}
\label{ABcosy}
\end{figure}

\section{Results and Discussion}
We investigate the effects of emotional speech intelligibility, the perception and rendering of single and mixed emotions, the effectiveness of emotion-guided prompt learning, and zero-shot synthesis of unseen emotional speech.
The subjective evaluation results are presented in Fig.~\ref{singleAB}, Fig.~\ref{ABmix}, Fig.~\ref{ABcosy}, Fig.~\ref{mosall}, Table~\ref{BWS}, Table~\ref{ourmos}, while the objective evaluation results are shown in Table~\ref{wer}. 

\subsection{Speech Intelligibility Evaluation}
For objective evaluations, we assess both single-emotion and mixed-emotion speech generation, as presented in Table~\ref{wer}. \textit{PUE} demonstrates superior performance in mixed-emotion speech generation over the two strong baselines. These results highlight the model’s effectiveness in producing clearer emotional speech over the state-of-the-art while preserving the linguistic content for mixed-emotion speech generation.

\subsection{Single Emotion Generation}

For objective evaluations, Table~\ref{wer} further shows that the proposed \textit{PUE} model consistently outperforms all baseline systems in terms of speech intelligibility for single-emotion speech generation across various emotional categories in the original test set, confirming its effectiveness in capturing single-emotion characteristics for emotional TTS.

For subjective evaluations, Fig.~\ref{singleAB} further indicates that our proposed \textit{PUE} outperforms the mix baseline and the cosy baseline substantially in terms of perceived speech quality for single-emotion generation, which further verifying its effectiveness in modeling and synthesizing high-quality single-emotion speech.

\subsection{Mixed Emotion Rendering}
We conduct both objective and subjective evaluations to assess the effectiveness of mixed emotion rendering.
\subsubsection{Objective Evaluation}
From an objective evaluation perspective (Table~\ref{wer}), \textit{PUE} surpasses all strong baselines for all mixed emotion scenarios in speech intelligibility according to the WER results, demonstrating its ability to generate more intelligent mixed-emotion speech.
\subsubsection{Subjective Evaluation}
For the subjective evaluation using Best-Worst Scaling (BWS) as presented in Table~\ref{BWS}, 94.44\% of listeners identified the mixed-emotion speech containing 90\% anger as expressing the strongest sense of outrage, compared to versions with lower anger proportions (30\% and 60\%). Moreover, 80.56\% of participants perceived the non-mixed anger utterance as the weakest expression of outrage. These findings are consistent with the hypothesis that increasing the proportion of anger enhances the perceived intensity of the target emotion, thereby demonstrating the controllability of mixed emotion rendering through proportion adjustment. A small proportion of listeners (5.56\%) selected the inverse ranking, likely due to inter-listener variability or the inherent subjectivity of emotional perception.

Furthermore, to investigate the impact of emotion-relevance parameter settings, we conduct subjective evaluations -MOS- under varying configurations, as shown in Table~\ref{ourmos}. The results indicate that the highest speech quality is achieved when surprise is mixed with 90\% of another emotion, outperforming lower mixing ratios. This finding further supports that increasing the proportion of secondary emotions enhances the perception of the target emotion, demonstrating the effectiveness of \textit{PUE} in modeling mixed emotions through flexible emotion-relevance parameter design.

\begin{table}[t!]
\centering
\caption{MOS results comparison with 95\% confidence interval of the proposed \textit{PUE} under different test cases: original test database; 100\% surprise mixed with varying proportions of other emotions: 30\%, 60\%, and 90\% for angry, sad, or happy.}
\begin{tabular}{l l}
\toprule
\textbf{Prompt unseen emotions} & \textbf{MOS} \\\midrule
Mix surprise with 30\% angry/sad/happy & 3.12$\pm$0.16 \\
Mix surprise with 60\% angry/sad/happy & 2.50$\pm$0.13 \\
Mix surprise with 90\% angry/sad/happy & \textbf{3.57$\pm$0.18} \\
\bottomrule
\end{tabular}
\label{ourmos}
\end{table}

\subsection{Effectiveness of Emotion-guided Prompt Learning}

To evaluate the perceived speech quality of the generated mixed-emotion speech, we further conduct AB preference tests comparing our proposed \textit{PUE} model against existing baseline systems, as illustrated in Fig.~\ref{ABmix} and Fig.~\ref{ABcosy}. 
\subsubsection{\textit{PUE} vs. mix}
Specifically, Fig.~\ref{ABmix} presents the results of AB preference tests in which listeners were asked to compare speech samples generated by the proposed \textit{PUE} model against those produced by the mix baseline under multiple mixed-emotion configurations. In this evaluation, the emotion surprise was blended with three different emotions—angry, sad, and happy—at varying proportions of 30\%, 60\%, and 90\%. This setup allows for a comprehensive investigation of the proposed model’s capability to generate emotionally rich and perceptually convincing speech across a spectrum of emotion combinations.

The results demonstrate a clear preference for the \textit{PUE} model across all tested configurations. When surprise was mixed with happy, a substantial 87.04\% of listeners preferred the speech generated by our model, while only 9.26\% favored the mix baseline. Similarly, for the blend of surprise and angry, 74.07\% of participants selected the \textit{PUE} model, compared to 25.93\% who preferred the baseline. In the case of surprise mixed with sad, our model was favored by 68.52\% of listeners, while 31.48\% selected the baseline. Notably, among the three emotion combinations examined, the surprise + happy pairing yielded the highest listener preference, indicating that this blend may be more perceptually congruent and emotionally harmonious.

A plausible explanation for this outcome is that surprise and happy share several overlapping acoustic features—such as elevated pitch contours, faster articulation rates, and higher vocal energy—that may result in more seamless emotional integration. These shared prosodic traits likely contribute to a more fluid and natural-sounding synthesis, making the resulting speech appear more expressive and emotionally engaging. This acoustic alignment facilitates the generation of emotionally blended speech that resonates more effectively with listeners, thereby enhancing the overall perceived quality and authenticity of the synthesized output.

These consistent listener preferences strongly indicate the effectiveness of the proposed emotion-guided prompt learning mechanism in generating perceptually superior emotional speech. The notable preference margins across different emotion combinations highlight the model’s effectiveness and flexibility in capturing subtle affective nuances introduced by varying the mixture ratios. Compared to the baseline system, which relies on conventional emotion mixing strategies, the \textit{PUE} model demonstrates a more nuanced understanding of emotional context, leading to more natural and emotionally appropriate speech synthesis.

Overall, the AB preference results in Fig.~\ref{ABmix} validate the superiority of the \textit{PUE} model over existing approaches, reinforcing the advantage of using prompt-based conditioning for fine-grained emotion control in mixed-emotion speech generation.

\begin{figure}[t]
\centering
\includegraphics[width=79mm]{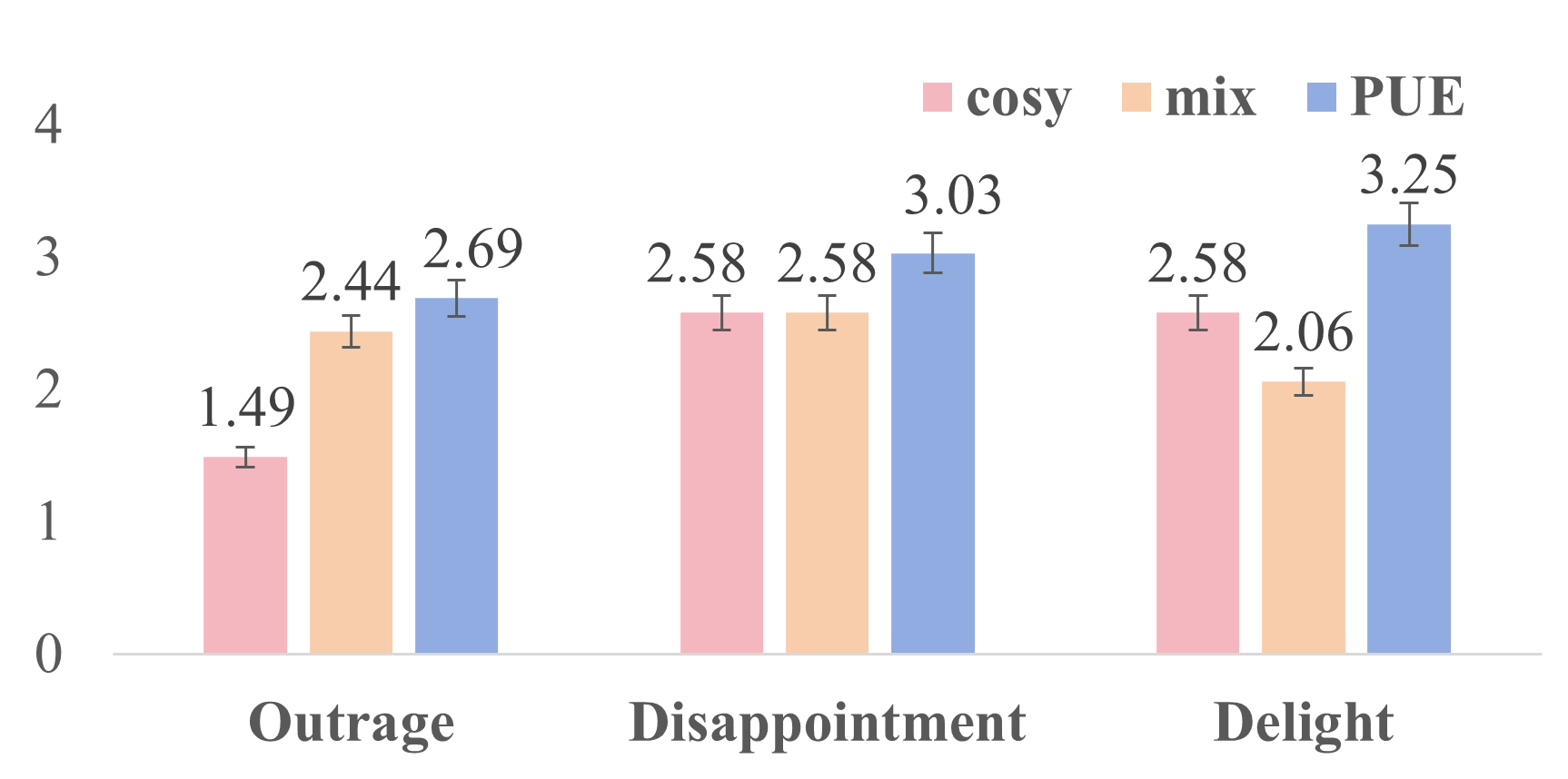}
\caption{Comparison of subjective evaluation results for MOS tests with 95\% confidence interval across cosy, mix, and the proposed \textit{PUE} models on three unseen emotions: outrage (surprise + 90\% angry), disappointment (surprise + 90\% sad), and delight (surprise + 90\% happy).}
\label{mosall}
\end{figure}
\subsubsection{\textit{PUE} vs. cosy}
Fig.~\ref{ABcosy} presents the results of AB preference tests comparing the proposed \textit{PUE} model against the \textit{cosy} baseline across three mixed-emotion conditions: surprise + angry, surprise + happy, and surprise + sad. The results show a consistent and substantial preference for the \textit{PUE} model across all emotion combinations. 
These settings were designed to assess the speech quality of the \textit{PUE} model in generating perceptually coherent and emotionally appropriate speech across a range of emotion mixtures.
Specifically, 74.07\% of listeners preferred our model over the cosy baseline (25.93\%) for the surprise + angry blend. For surprise + happy, the preference was even more pronounced, with 81.48\% of participants selecting our model, compared to only 11.11\% for the baseline. In the case of surprise + sad, 68.52\% favored the \textit{PUE} model, while 22.22\% chose the cosy baseline.

These results provide strong empirical support for the superiority of the emotion-guided prompt learning mechanism in \textit{PUE}, which facilitates more nuanced and coherent mixed-emotion synthesis compared to the cosy approach. Notably, the surprise + happy combination again achieved the highest listener preference, reinforcing earlier findings from comparisons with the mix baseline. This suggests that the acoustic and prosodic compatibility between surprise and happy—such as similar pitch dynamics and energetic vocal expression—contributes to a more perceptually pleasing and emotionally expressive synthesis.

Overall, the consistent performance gains across all mixed emotion conditions in Fig.~\ref{ABcosy} highlight the effectiveness and generalizability of the \textit{PUE} model. By leveraging emotion-conditioned prompts, our approach effectively models the interplay between different emotions and yields speech that is more natural, emotionally rich, and preferred by human listeners over existing baseline systems.

\subsection{Zero-shot Unseen Emotional Speech Synthesis}
To further evaluate the unseen emotion generation capabilities of our proposed \textit{PUE} model, we conduct subjective evaluations for zero-shot unseen emotional speech synthesis, focusing on three novel composite emotions not present during training: outrage (surprise + 90\% angry), disappointment (surprise + 90\% sad), and delight (surprise + 90\% happy). The subjective evaluation results are presented in Fig.~\ref{mosall}, which illustrates the MOS with 95\% confidence intervals for the three emotion categories, comparing the \textit{PUE} model against both the cosy and mix baselines. The decision to use a 90\% proportion for the secondary emotion in each blend is based on prior findings in Table.~\ref{wer} and Table.~\ref{BWS}, which demonstrate that higher proportions led to better perceptual outcomes than lower mixing ratios such as 30\% or 60\%. This setup ensures a stronger emotional rendering in the synthesized speech and allows for clearer listener perception of the intended emotional state.

From the results in Fig.~\ref{mosall}, we observe that the \textit{PUE} model consistently achieves the highest MOS across all three unseen emotion types, significantly outperforming both the cosy and mix baselines. This performance gap demonstrates the superior capability of our model in generating high-quality and emotionally coherent speech for emotion blends it has never encountered during training. In particular, the \textit{PUE} model exhibits notable improvements in perceived speech quality for mixed emotional expressiveness, underscoring its zero-shot generalization strength.
The strong performance across these unseen composite emotions also reflects the flexibility and compositional nature of the emotion-guided prompt learning mechanism. By explicitly modeling emotion relevance through prompt parameters, the \textit{PUE} model is able to extrapolate emotional characteristics from known emotions and synthesize novel emotional states in a controllable and natural-sounding manner.
A demo page with synthesized emotional speech samples is available at the provided link \footnote{\url{https://anonymous.4open.science/w/PUE-785D}}.

\section{Conclusion}
This paper presents an innovative \textit{PUE} approach for expressive speech synthesis through emotion-guided prompt learning on an LLM-TTS architecture. The proposed \textit{PUE} significantly advances expressive TTS systems by enabling the generation of unseen emotional speech through modifying categorical emotion-relevant prompts, eliminating the need
for additional training on new emotion categories. 
Both subjective and objective evaluations demonstrate that \textit{PUE} outperforms state-of-the-art baselines, demonstrating its effectiveness in capturing diverse emotional nuances and facilitating zero-shot emotion generation.
The code will be made available upon acceptance for the research community.

\bibliographystyle{IEEEtran}
\bibliography{mybib}

\end{document}